\documentclass{www2009-submission}

\usepackage{url}

\newtheorem{theorem}{Theorem}[section]

\newtheorem{lemma}[theorem]{Lemma}

\numberwithin{equation}{section}

\newcommand{\ignore}[1]{}
\newcommand{\shortvers}[1]{}

\begin{document}
\title{Bid Optimization for Broad Match Ad Auctions}

\author{
Eyal Even Dar\thanks{Google Research, 76 9th Ave,
New York, NY 10011, \{evendar,mirrokni,muthu\}@google.com }
\and Yishay Mansour \thanks{Google Research and Tel-aviv University, 76 9th Ave,
New York, NY 10011, mansour@google.com }\and
Vahab S. Mirrokni\footnotemark[1]
\and S. Muthukrishnan\footnotemark[1]
\and Uri Nadav \thanks{Tel-aviv University, Tel-Aviv, 69978, urinadav@gmail.com}}

\maketitle
\begin{abstract}
Ad auctions in sponsored search support ``broad match'' that 
allows an advertiser to target a large number of queries while bidding only 
on a limited number. 
While giving more expressiveness to advertisers, this feature  makes it challenging
to optimize bids to maximize their returns:
choosing to bid on a query as a broad match because it provides 
high profit results in one bidding for related queries which may yield low or even
negative profits.

We abstract and study the complexity of the {\em bid optimization problem} which is to 
determine an advertiser's bids on 
a subset of  keywords (possibly using broad match) so that her profit is maximized.
In the query language model when the advertiser is allowed to bid on all queries as broad match, we present
an linear programming (LP)-based polynomial-time algorithm that gets the optimal profit.
In the model in which an advertiser can only bid on keywords, ie.,  
a subset of keywords as an exact or broad match, we show that this problem is not approximable
within any reasonable approximation factor unless P=NP. To deal with this hardness result, 
we present a constant-factor approximation when the optimal profit significantly 
exceeds the cost. This algorithm is based on rounding
a natural LP formulation of the problem.
Finally, we study a budgeted variant of the problem, and show that in the query language model, one
can find two budget constrained ad campaigns in polynomial time that implement the optimal bidding strategy.
Our results are the first to address bid optimization under the broad match feature which is
common in ad auctions. 
\end{abstract}




\keywords{Sponsored Search, Ad Auctions, Optimal bidding, Bid Optimization}

\section{Introduction}
Sponsored search is a large and thriving market with three distinct players. 
{\em Users} go to search engines such as Yahoo! or Google and pose queries; in the process,
they express their intention and preferences. {\em Advertisers} seek to place advertisements 
and target them to users' intentions as expressed by their queries. Finally, {\em search engines}
provide a suitable mechanism for doing this. Currently, the mechanism relies on having advertisers bid on the 
search issued by the user, and 
the search engine to run an {\em auction} at the time the user poses the query to determine the 
advertisements that will be shown to the user. As is standard, the advertiser only pays if the 
user clicks on their ad (the "pay-per-click" model), and the amount they pay is determined by 
the auction mechanism, but will be no larger than their bid. 

In this paper, we assume the perspective of the advertiser. The advertisers need to target their ad campaigns to
users' queries. Thus, they need to determine the set $S$ of queries of their interest. Once that is determined,
they need to strategize in the auction that takes place for each of the queries in $S$.
A lot of research has focused on the game theory and optimization 
behind these auctions, both from the search engine~\cite{AGM,Varian,EOS,AE,MSVV,BCIMS} and advertiser~\cite{etesami,FMPS,Deep,mps} points of view. 
There has been relatively little prior research on how advertisers target their campaign, i.e., how they
determine the set $S$.

The criterion for choosing $S$ is for the advertiser to pick a set of {\em keyphrases} that 
searchers may use in their query when looking for their products. The central challenge then is to 
match the advertisers keyphrases with the potential queries issued by the users. It is difficult if not 
impossible for the advertisers to identify all possible variations of keyphrases that a user looking for 
their product may use in their query. As an example, consider a vendor who chooses the keyphrase 
{\em tennis shoes}. Users searching for them may use singular or plural, synonyms and other variations (``clay court footwear''),  
may misspell (``tenis shoe''), use extensions (``white tennis shoes'') or reorder the words (``shoes lawn tennis''). In fact, 
users may even search using words not found in the keyphrase (``Wimbledon gear'', ''US Open Shoes'', ``hard court soles''), 
and may still be of interest to the advertiser. These artifacts such as plurals, synonyms, misspellings, extensions, and 
reorderings are very common, and the problems get compounded since typical ad campaigns comprise several keyphrases, each 
with its own set of artifacts. 

Major search engines help advertisers address this challenge by providing a structured bidding language. 
While the specific details differ from search engine to search engine~\cite{msn,yahoo,google}, at the highest level, 
the bidding language supports two {\em match types}: exact and broad. In {\em exact} matchtype (called ``exact''  
in MSN AdCenter and Google, and ``standard'' in Yahoo), 
ad would be eligible to appear when a user searches for the specific keyphrase without any other terms in the query, 
and words in the keyphrase need to appear in that order. In {\em broad} matchtype (called ``broad'' in MSN, related 
to ``phrase'' and ``broad'' in Google, and ``advanced match type'' in Yahoo), the system automatically makes 
advertisers eligible on relevant variations of their keyphrases including for the various artifacts listed earlier, even if the 
search terms are not in the keyphrase lists. Thus, the search engines automate the aspect of detecting artifacts and matching the query to
keyphrases of interest to advertisers.\footnote{These match types may be further modified by ensuring that the ad be 
{\em not} shown on occurrence of certain keywords in the query; this feature (called ``negative'' in MSN and Google or 
"excluded'' by Yahoo) and other targeting criteria associated with keyphrase campaigns do not change the discussion 
and the results here.} Thus the task of advertisers becomes determining the keyphrases and choosing the match type on each.

The question we address here is, how does an advertiser bid in presence of these match types? 
Say each query $q$ has a value $v(q)$ per click for the advertiser that is known to the advertiser and is
private. Further,
 we let $c(q)$ be the expected price per click 
and let $n(q)$ be the expected number of clicks. These are statistical estimates provided by the 
search engines~\cite{Googlestats,Yahoostats,MSNstats}. Then, we consider two optimization problems: (i) 
in one variant, we assume that the advertiser wishes to maximize their {\em expected profit}, that is, $\sum_q (v(q)-c(q))n(q)$,
and (ii) in the other variant, given a budget $B$ for the advertiser, we assume that the 
advertiser wishes to maximize their {\em expected value}, that is, $\sum_q v(q)n(q)$ subject to the condition that 
the expected spend $\sum_q c(q)n(q)$ does not exceed the budget. 

The technical challenge arises due to {\em query dependencies}. When one bids on a keyphrase for query 
$q$, as a result of a  broad match, it may apply to query $q'$ as well. The advertiser has different values
$v(q)$ and $v(q')$ on these because users for $q$ and $q'$ differ on their intentions and therefore on their 
respective values to the advertiser. So, the advertiser may make good profit on $q$ and may wish to bid on that query, 
but is then forced to implicitly bid on $q'$ as well, and may even make negative profit on $q'$!
Under what circumstances is it now desirable for the advertiser to bid for $q$?

Note that query dependence is a fundamental aspect of sponsored search since advertisers can 
realistically only choose and strategize on a small set of keyphrases because of the effort involved, and 
have to typically rely on the search engine to carefully apply their strategy to variants of their keyphrases.
But beyond that, even an ad campaign that is willing to exert a lot of effort and use
a large number of keyphrases or relies on a search engine to provide rich bidding languages~\cite{Johannes}
 will still find it impossible to include all 
search variations of the keyphrases
as exact matches, and must necessarily rely on broad match for the variations that search users develop and prefer
over time. Thus, the advertisers bid implicitly on queries on which they can not directly control the tradeoff
between the cost and the value. 

Query dependence introduces a complex optimization problem of trading off the benefits of bidding on a keyphrase against the impact of bidding on its dependent queries.
In the sponsored search world, there is a keen awareness of this complexity of bidding, and most search engines and  third-party bidding agents provide
detailed tips and guidelines for advertisers~\cite{yahoohandbook,copernic}. 
Beyond these guidelines, what is missing is a clear
theoretical understanding of the tradeoffs and the complexity of the bidding problem that advertisers face.

We initiate principled study of bidding in presence of broad matches.
Specifically, our contributions are as follows.
\begin{enumerate}
\item
We abstract two models --- query and keyword language models --- to study bidding optimization problems. 

In the query 
language model, the advertiser bids directly on user queries and wishes to determine which query if any to bid on, 
to maximize expected profit. This models both the theoretical extreme where an advertiser can bid on any of the queries the 
search engine will see, and the practical reality where the advertiser has a select set of queries in mind and wishes only 
to optimize within that set. 
In the keyword language model, advertisers may bid only on a subset of queries, and broad match 
implicitly derives bids as needed. This directly models the common reality.

\item
We present efficient, polynomial time algorithms for the bid optimization problem under these two models. 

In query bidding, we get a polynomial-time algorithm that maximizes the profit, using a reduction to the well-known 
Min-Cut problem in graphs. This is in contrast to the poor performance of natural greedy algorithms for this problem.
We also study the budgeted variant of the problem, and propose a novel strategy using {\em two} distinct budgeted ad campaign that gets the optimal profit. We do so by studying the structure of the basic feasible solutions of a corresponding linear programming formulation of the problem.

For keyword bidding, we show that even limited instances are NP-Hard to not only optimize,
but even to approximate;  
to deal with this hardness result, we present a constant-factor approximation when advertisers profit
following an  optimal bid is considerably greater than her cost. This result is based on applying a
randomized rounding method on the optimal fractional solutions of the linear programming relaxation of the
problem. 
\end{enumerate}

These represent the first known theoretical results for the problem of bid optimization 
in presence of broad matches, a problem advertisers face now since this feature is offered by the major search engines. 
Prior research in bid optimization for advertisers~\cite{etesami,Deep,David} primarily focused on determining suitable bids for 
exact match types and does  not study the query dependence and implicit bids; ~\cite{FMPS,mps} studied 
the problem of maximizing the number of clicks, and not the profit which is the more standard metric.
At the technical core, our challenge is to tradeoff positive 
profit from bidding on a keyphrase that applies to one query $q$ against possibly a negative profit from 
the implied bids of broad match on queries $q'$. This query dependence is a novel feature in sponsored search auctions, not
explicitly studied in prior literature,
and our results for this problem may have applications beyond, in the general auction theory area. 

Finally, we  report experimental results on a small family of instances of the bid optimizations problem, 
and compute the optimal bidding using the integer linear programming formulation. Our main observation in these experiments is that by considering only the broad match, we do not lose much in the maximum profit of the solution. This supports
our hope that under reasonable circumstances (similar to the ones in our experiments), considering only broad match is effective, and in turn, that would enable advertisers to focus on campaigns with small lists of keyphrases.

\section{Model}

We consider the optimization problems that an advertiser faces while
bidding in an auction for queries with a broad match feature.


\bigskip
\noindent{\bf The Advertiser.} We consider a single advertiser who
is interested in showing her ad to users after they search for
queries from a set $Q$. The advertiser has some utility from having
a user click on her ad. In reality, clicks associated with different
queries may bring have utility to the advertiser; The advertiser has
a value of $v(q)$ units of monetary value associated with a `click'
that follows a query $q\in Q$.

We assume a posted price model where prices are posted and the
search volume of every query as well as its click through rate
(i.e., the probability that users would click her ad) are known to
the advertiser. Namely, every query $q$ is associated with a pair of
parameters, known to the advertiser, $(c(q), n(q))$, where $c(q)$ is
the per click cost of $q$, and $n(q)$ is the expected number of
clicks that would result from winning  $q$ (the expected number of
clicks can be determined from the search volume of $q$ and the
advertiser's specific click through rate for $q$).

Thus, when an advertiser wins a query $q$, her overall profit~\footnote{In this
paper, we use terms utility and profit interchangably.} from
winning, denoted $w(q)$ is
$$w(q) = \left(v(q)-c(q)\right)n(q) ~.$$
Note that although each query has a positive value, winning it may
result an overall negative profit.

\bigskip

\noindent{\bf Bidding languages.} A bidding language is a way for an
advertiser to specify her value or willingness to pay for
queries. Eventually, the auctioneer needs to have a bid for every
possible query \footnote{A  bid of $0$ for a query may be regarded as
the default in a case where the advertiser is not explicitly
interested in a query $q$ and nor in queries that $q$ match broadly.}. 
The choice of a bidding language is critical for the auction
mechanism. At the one extreme, it may be infeasible to allow an
advertiser to specify explicitly her value for every possible query.
On the other hand, a language that is too restrictive would not
allow an advertiser to communicate her preferences properly.

In order to study the complexity of the optimal bidding in the broad
match framework while taking into account the intersections among
broad matches for different keywords, we first consider a bidding
language in which an advertiser can specify a bid for every query
$q$ but only as a broad-match. We refer to this language as the {\em
query language}.

To allow the most accurate description of an advertisers value per
query, the ultimate way is to let the advertiser specify all
possible queries with exact or broad match, and a monetary bid for
each of them. If an advertiser is allowed to bid on each type of
query as an exact match as well as broad match, she can decide for
each query independent of the other queries, and the complexity of
the bidding problem is not captured in such bidding language.

To capture the complexity of the optimal bidding problem and the
fact that advertisers may only bid on a subset of queries, we study
the \textit{keyword language} that allows advertisers to place a bid
only on (single) keywords or short phrases. More precisely, in the
keyword language, we assume  that advertisers are allowed to bid
only on a subset $S\subset Q$ of queries.

A further improvement of this language would allow the advertiser to
specify, besides a value bid for $s\in S$,  whether $s$ is to be
matched exactly or broadly.

A bid $b\in \mathbb{R}_+^{|Q|}$ in some bidding language is
associated with a set of `winning queries' denoted by $\varphi(b) =
\{q\in Q ~|~ b(q)\geq c(q)\}$. A subset $T$ of queries which is a
winning set of some bid $b$ is referred to as a \emph{feasible
winning set}. The utility associated with a winning set $T$ is
$$u(T)=\sum_{q\in T} \left(v(q)-c(q)\right)n(q),$$ where $v(\cdot)$
and $n(\cdot)$ are advertiser specific.

A feasible winning set with optimal utility is referred to as an
optimal winning set.

\bigskip \noindent{\bf The Auction.} For every query, the auctioneer should decide the bid
of every advertiser. This decision is easy for queries on which the
advertiser bids explicitly (as an exact match). However, for the
queries that the advertiser has not bid directly, but only through
a broad match framework, the auctioneer should compute an
appropriate bid for the advertiser to participate in the auction.

A natural way for setting such a value is to aggregate the bid
values of all the phrases matched by the query. While there are
several choices for the aggregation method, in this paper, we
consider the $\max$ aggregation operator --- when a query $q$
matches phrases $w_1,\ldots,w_k$ from the advertiser list of
phrases, its bid is interpreted as $b(q) = \max_{i}{b(w_i)}$. 

We can now state formally the bid optimization problem. Given
advertiser's specific data (A set $Q$, value for queries $v$, search
volume and click through rates $n(\cdot)$ ) and a bidding language
$\mathcal{L}$, an optimal bid $b^*$, is a feasible bid in the
language $\mathcal{L}$ that maximizes the advertisers' utility from
winning a set $\varphi(b)$ of queries. Formally,

\begin{equation}
  b^* \in argmax_{b \in\mathcal{L}} \{ u(\varphi(b)) \} .
\end{equation}

\bigskip\noindent{\bf Query dependencies.}
We say that a query $q$ depends on a query $q'$ if winning query
$q'$ implies winning query $q$. In the broad match auction in which
the bid interpretation strategy is done using the max operator, this
happens if $q$ matches $q'$ broadly, and its cost $c(q)$ is less
than that of $c(q')$. In other words,  if a bid $b$ wins $q'$, it
must be that $b(q')\geq c(q')$, but the interpreted bid for $q$ is
then at least $b(q')\geq c(q)$ since $c(q')\ge c(q)$, hence the bid
$b$ must be winning $q$ as well. As a result, the cost structure
incurs a set of pairs $(q',q)$ where the first entry of each pair
$q'\in S$ is a valid phrase in the bidding language and the second
entry is a valid query in the set of queries $Q$ such that winning
query $q'$ implies winning query $q$. This set of pairs is denoted
by $\cal C$ and formally:

$${\cal C} = \{ (q', q)\vert q'\in S, q\in Q, q \mbox{\ matches\ } q' \mbox{\ broadly\ }, c(q')\ge c(q)\}.$$

Moreover, we define $D(q) = \{q'\vert (q', q)\in {\cal C}\}$, and
$N(q) = \{q'\vert (q, q')\in {\cal C}\}$.

\bigskip {\bf \noindent Budget-constrained Ad Campaigns.} A variant of
the optimal bidding problem in the broad match framework is to find
a set of queries to bid on that maximizes the total value of the
queries won by the advertiser subject to a budget constraint, i.e,
our goal is to bid on a subset $T$ of queries to maximize
$\sum_{q\in T} v(q) n(q)$ subject to the budget constraint
$\sum_{q\in T} c(q) n(q) \le B$. To handle such a budget constraint,
we assume that one can run a {\em budget-constrained ad campaign} by
bidding on a subset $T$ of keywords and setting a budget $B$.
Assuming $B'=\sum_{q\in T} c(q) n(q)$, there are two possibilities
in this budget-constrained ad campaign: (i) If $B'\le B$, the
auction is run in a normal way and the value from this ad campaign
for the advertiser is $\sum_{q\in T} v(q)n(q)$, (ii) On the other
hand, if $B'> B$, we assume that the queries arrive at the same rate
and as a result, for each query, we get $B'\over B$ fraction of the
value of an ad campaign without the budget constraint. In other
words, the value that the advertiser gets is $\sum_{q\in T} v(q)
n(q) {B'\over B}$. We can also interpret the above assumption by a
throttling method in which, in order to cope with the budget
constraint, at each step, we let the advertiser participate in the
auction with probability $B'\over B$.


\section{Bidding in the query language}
\label{sec:querylang}

In this section, we study the query language that allows placing a
bid on every query.
\ignore{
 but only in a broad-match fashion; specifying
exact match while bidding on a query is not allowed. 
The advantage
of studying this model is two-fold: it helps understanding the
complexity of broad match versus exact match bidding problems, and in
particular, it shows the advantage of using linear programming
formulation for this problem compared to greedy algorithms.
In addition to this advantage, by solving the optimal broad match solution
in the query language, we might be able to bid on a small number
of keywords as a broad match, and get a comparable performance
with the case that we bid on a large number of queries as exact match.
The main advantage of smaller ad campaigns is that advertisers deal with 
smaller number of queries.
We observe this advantage experimentally in our simulation 
results that we present in Section~\ref{sec:simulation}.
}
We observe that in the query language, the task of computing an
optimal bid is equivalent to that of computing an optimal winning
set: Given an optimal feasible set $T$ set a bid $b(q)=c(q)$ for
every query $q\in T$ with positive weight and $b(q)=0$ otherwise.

\begin{lemma} A bid $b$ derived from an optimal winning set $T$, as
described above, is an optimal bid.
 \end{lemma}

\begin{proof}
By construction, the bid $b$ wins all the queries with positive
weight from $T$, and every other query must belong to $T$ (otherwise
$T$ would not be feasible).
\end{proof}

We therefore consider algorithms for computing an optimal feasible
winning set. First, we consider a greedy algorithm, denoted by \emph{Max-Margin Greedy}.
Initially,  \emph{Max-Margin Greedy} sets the winning set to be empty. Then,
iteratively, it adds a bid on a query with the highest marginal
benefit to the winning set utility. Unfortunately, \emph{Max-Margin Greedy}
fails to compute an optimal winning set due to the following
example.

\bigskip
\noindent{\bf Example.} Consider a set $Q$ of queries which contains
$n$ keywords and another $n\choose 2$ queries, each of which is a
pair of keywords. The cost of each query is set to $\$1$. Hence, the
query dependencies is such that winning a keyword implies winning the
set of $n-1$ queries made of pairs of keywords in which this
keyword appears. The value of a keyword is set to \$2; The value of
a each pair is set to $1-1.5/n$. So, every keyword attains a
positive utility of \$1, and every pair causes a loss of $\$1.5/n$.
Initially, \emph{Max-Margin Greedy}'s bid is empty. At this point \emph{Max-Margin Greedy}
is stuck --- every single query it adds to the winning set results
in a negative overall utility. Thus, this instance, {\em Max-Margin Greedy}
would yield 0 utility. An optimal solution wins all the queries and
has a utility of $n\times(2-1) - {n\choose 2}(1-\frac{1.5}{n} -1) =
\frac{n-3}{4}$.

One can explore other variants of greedy algorithms for this problem. For example,
a natural greedy algorithm is {\em Max-Rate Greedy} algorithm: 
Initially set the winning set to the empty set, and then iteratively, add a bid on a query with the 
highest ratio of marginal profit over the marginal cost, or the query with the
highest ratio of marginal value over marginal cost. We note
that all these iterative greedy algorithms pefrom poorly for the above example. 
Even a significant look-ahead will not resolve this bad example.

We turn to the next algorithm \emph{OptBid1} for computing an
optimal winning set. \emph{OptBid1} is a solution to
the following integer linear program:

\begin{eqnarray}\mathrm{ILP}&:& \max \sum_{q_i\in Q} X_{q_i} w(q_i) \nonumber\\
 {\text For \ every \ pair \ }(q_j,q_i)\in \mathcal{C} &:& X_{q_i}-X_{q_j}\geq 0 \nonumber \\
 \forall q_i\in Q &:& X_{q_i}\in\{0,1\} 
 \label{IP:ineq:pair}
\end{eqnarray}

For every query $q$, an integral variable $X_q$ is a 0-1 variable
which is equal to 1 if and only if $q$ belongs to the winning set of
queries. In order to solve the above ILP, we relax it to a linear
program where instead of integer 0-1 variables, we have fractional $X_q$
variables with values between 0 and 1 ($0\le X_q\le 1$).
Here, we observe that the integrality gap of this linear programming relaxation is 1,
i.e., for any instance of this linear program, there exists
an optimal solution $X^*$ in which all the values are integer $X^*_q\in \{0, 1\}$
for all $q\in Q$.

\begin{lemma}
The integrality gap of the linear programming relaxation of the ILP~\ref{IP:ineq:pair} is one.
\end{lemma}
\begin{proof}
The lemma follows from the fact the the constraint matrix of the LP
relaxation of ILP~\ref{IP:ineq:pair} is totally
uni-modular\footnote{A matrix $A$ is totally uni-modular if every
square submatrix of it is uni-modular, i.e., every submatrix has a
determinant of 0, -1 or +1.}. A sufficient condition for a matrix to
be totally uni-modular is that every row has either two non-zero
entries, one is $1$ and the other $-1$, or a single non zero entry
with value $1$ or $-1$. An integer program whose constraint matrix
is totally uni-modular and whose right hand side is integer can be
solved by linear programming since all its basic feasible solutions
are integer (see \cite{papadimitriou} pp. 316).
\end{proof}

The above lemma implies the following polynomial-time algorithm
{\em OptBid1} for optimal bidding in the query language: compute a basic
feasible solution $X^*$ of the LP relaxation
of ILP~\ref{IP:ineq:pair}, and find a bidding strategy corresponding
to the winning set of $X^*$, i.e., $\{q\in Q\vert X^*_q = 1\}$.

\begin{figure}\label{fig:OptBid2}
\centering
\epsfig{file=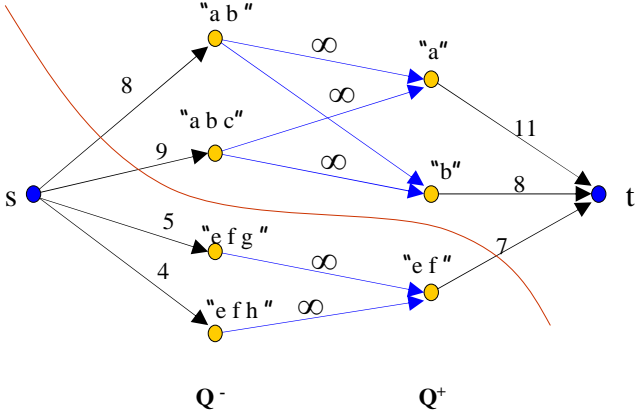}
\caption{An example of running algorithm OptBid2 on the set of queries
(with the following profit: $\{(a,11),(b,8),(ab,-8),(abc,-9),(ef,7),(efg,-5),(efh,-4)\}$), 
and dependency graph as illustrated. ObtBid2 will choose the winning set $\{a,b,ab,abc\}$. The optimal bid is $\{(a,11),(b,8)\}$.}
\end{figure}

The running time of Algorithm \emph{OptBid1} is that of solving a
linear program with $\Omega(|Q|^2)$ constraints, which although
polynomial in $|Q|$, might be inefficient. Next, we present a
faster algorithm, \emph{OptBid2}.

For the purpose of presenting Algorithm \emph{OptBid2}, we define a
weighted flow graph $G=(V,E)$, derived from the input. The vertex
set of $G$ is $V=\{s,t\}\cup Q^+\cup Q^-$, where $s$ is a source
node,  $t$ is a target node and $Q^+$ and $ Q^-$ are the sets of
queries with positive/non-positive weights respectively, {\emph
i.e.}, $Q^+ \equiv\{q~|~ w(q)>0\}$.
The source vertex $s$ is connected to each vertex $q\in Q^-$ with an
edge of weight $|w( q)|=|(v(q)-c(q))n(q)|$. The target vertex $t$ is
connected with each vertex $p\in Q^+$ with an edge of weight $w(p)$.
Two vertices $q\in Q^-, p\in Q^+$ are connected with an edge of
weight $\infty$ if and only if $(p,q)\in \mathcal{C}$.

\bigskip\noindent{\bf Algorithm \emph{OptBid2}}
\begin{enumerate}\item Compute a min-cut of $G$. Let $S,T$ be the
two sides of the cut.

\item Assume, without loss of generality, that $t\in
T$. Return $T\setminus \{t\}$, that is, the set of queries that are
on the same side of the cut as $t$ is an optimal winning set.
\end{enumerate}

The running time of Algorithm \emph{OptBid2} is that of min-cut,
{\emph i.e.}, $O(|Q|^3)$~\cite{SW97}.

\begin{theorem}
Algorithm \emph{OptBid2} finds an optimal winning set.
\end{theorem}
\begin{proof}
We show $T$ is an optimal winning set using the dual program of ILP.
\begin{eqnarray*}\mathrm{DUAL}&:& \min \sum_{q\in Q^+} Z_{q} \\
 \forall q\in Q^+  &:& \sum_{q': (q,q')\in \mathcal{C}} Y_{q,q'} + Z_q\geq w(q) \\
 \forall q'\in Q^-&:& \sum_{q:(q,q')\in \mathcal{C}} Y_{q,q'}\leq -w_{q'} \ {\rm (Notice \ that \ } w_{q'}\leq 0)\\
 \forall (q,q')\in \mathcal{C} &:& Y_{q,q'}\geq 0\\
 \forall q\in Q^+ &:& Z_q\geq 0
\end{eqnarray*}
Let $f=(f_e)_{e\in E}$ be a maximum flow in $G$, with value $c$. For
every $(q,q')\in \mathcal{C}$, set $Y_{q,q'} := f_{q,q'}$ and for
every $q\in Q^+$, set $Z_q := w(q)- \sum_{q'|(q,q')\in \mathcal{C}}
Y_{q,q'}$. It is straightforward to verify that this is a feasible
solution of DUAL with value $\sum_{q\in Q^+} w(q) -c$.

Now, observe that $T$ is a feasible set in the query language. For
every pair $(q,q')\in \mathcal{C}$, we have that if $q\in T$ then
also $q'\in T$. Otherwise, the edge $(q,q')$, with weight $\infty$,
would be part of the cut. Thus, the value of the min cut is
$$c=\sum_{q\in Q^-\cap T} |w(q)| + \sum_{q\in Q^+\setminus T} w(q) .$$
and therefore, \begin{eqnarray*} u(T)&=&\sum_{q\in T} w(q) =
\sum_{q\in Q^+} w_q - \sum_{q\in Q^+\setminus T} w(q) + \sum_{q\in
Q^-\cap T} w(q) \\ &=& \sum_{q\in Q^+} w_q - c.\end{eqnarray*}
We already found a feasible solution for the dual of ILP, with the
same value. We therefore conclude, using the weak duality theorem,
that $T$ is an optimal solution of ILP.
\end{proof}

\section{Bidding in the keyword language}

In this section, we study optimal  bidding for 
the keyword language, where the advertiser is restricted to bid 
on a subset of (possibly short) queries $S\subset Q$.

Note that in the case that all queries have
positive utility, the optimal bid is trivial by
simply placing a high bid for every query in $S$. 
In addition, finding
the optimal bid when all queries are associated with a negative
utility is trivial (a bid of \$0 for every phrase in $S$ is
optimal). Moreover, in the case of uniform value from every query, 
the optimal bid is  easy --- a uniform bid
equal to the uniform value guarantees winning every query with
positive weight and losing every query with negative weight, which
is of course optimal.
In realistic settings, some queries have positive 
utility and some have negative utility. In this case the problem of
finding the optimal bid becomes intractable. More
precisely, as we show now, even when the set of
queries $Q$ is made up from single keywords and pairs of keywords, this 
problem becomes hard to approximate within a factor of 
$|S|^{1-\epsilon}$, for every $\epsilon>0$:

\begin{theorem}
In the keyword language broad match framework, it is NP-hard to approximate the optimal value of the optimal 
bidding problem  within a factor of $|S|^{1-\epsilon}$, for any
$\epsilon>0$.
\end{theorem}
\begin{proof}
We give a factor-preserving reduction to the independent set
problem. Given a graph $G$ with $n$ nodes, and $m$ edges, we
construct the following instance of our problem: put a singleton
keyword for each node $v$ of $G$ with weight $w_v = -(deg(v)-1)$,
and put a query consisting of a pair of keywords corresponding to each edge $e$ of G with weight
$1$. The maximum value we can get from picking a keyword is 1, and
we get this value if all of its neighbors do not appear in the
output. It can be seen that the optimum solution is an independent
set of nodes (since otherwise, we get zero or negative from a picked
node), and as a result, the maximum value is the same as the size of
the independent set.
\end{proof}

\subsection{A Constant-Factor Approximation}
In this section, in light of the above hardness result, 
we design a constant-factor 
approximation algorithm for a special case of the optimal bidding problem in the
keyword language in which the cost part of the optimal solution is
less than $1\over c$ of the value part of the optimal solution, for 
some constant $c>1$. Recall that $D(q) = \{ q'\vert q'\in S, (q',q)\in {\cal C}\}$.
Our algorithm is constant-factor approximation if
for any query $q\in Q\setminus S$, $\vert D(q)\vert$ is 
less than a constant $c'$.   
We present our result for the case that each query $q\in Q$ can be in the broad-match
set of at most two queries $q_1, q_2\in S$, i.e., $\vert D(q) \vert \le 2$. 
However, our result can be extended to more general 
settings in which query $q\in Q$ can be the broad match for a constant number of
queries $q_1, \ldots, q_{c'}\in S$ (for a constant $c'$).



 Based on the above discussion, 
we assume that $\vert D(q)\vert \le 2$ for any query $q\in Q$. Note that
the hardness result of the keyword language holds even for instances in which
$\vert D(q)\vert \le 2$ for all queries $q\in Q$. 
Also, let $E = \{ c(q)\vert q\in Q\}$.
Our algorithm is based on a linear programming relaxation of the optimal bidding problem
for the keyword language.   The integer linear program is as follows: 
\begin{eqnarray}
\text{ILP-Approx} \mbox{\ \ } \max & \sum_{s\in S}  (Z^{c(s)}_s+R_s) w_s \nonumber \\
  & + \sum_{q\in Q\setminus S} Y_q w_q \nonumber \\
  \forall q\in Q\setminus S, (s,q)\in {\cal C}, (r,q)\in {\cal C} & Y_q \leq Z^{c(q)}_s + Z^{c(q)}_r  \nonumber\\
  \forall q\in Q\setminus S, (s, q)\in {\cal C} & Y_q \ge Z^{c(q)}_s  \nonumber\\
  \forall s\in S, p,p'\in E & Z^p_s =\sum_{t\in E, t\le p} W^{t}_s  \nonumber \\
  \forall s\in S, p\in E& Z^p_s + R_s \le 1 \nonumber \\
  \forall q\in Q\setminus S & Y_q \in\{0,1\} \nonumber \\
  \forall s\in S & R_s \in\{0,1\} \nonumber \\
  \forall s\in S, p\in E & W^p_s, Z^p_s \in\{0,1\} \nonumber\\
\end{eqnarray}
Where the variables correspond to the following:
\begin{itemize}
\item $W^p_s$ for any $s\in S$ is the indicator variable corresponding to the bid of $p$ on query $s$ (as a broad match),  
\item $Z^p_s$ for any $s\in S$ is the indicator variable corresponding to the bid of at most $p$ on query $s$ (as a broad match),
\item $R_s$ for any $s\in S$ is the indicator variable corresponding to the exact match bid on query $s$,
\item $Y_q$ for any $q\in Q\setminus S$ is the indicator variable corresponding to winning query $q$ (as a result of bidding on queries in $S$).  
\end{itemize}

We relax the integer 0-1 variables in this integer linear program to fractional variables between zero and one, and then compute 
an optimal fractional solution for this LP relaxation. Then we round this fractional solution to construct a 
feasible (integral) bidding strategy. 

{\bf\noindent Rounding to an Integral Solution.} Given a fractional solution $(V, Z, W, Y)$ to the
LP, we round it to an integral solution $(V', W', Z', Y')$ as follows: 
for every query $s\in S$, we set $V'_s=1$ with probability 1, and $V'_s=0$ otherwise.
If $V'_s = 1$, we set $W'^p_s=0$ for all $p\in E$. Otherwise, for each $s\in S$, 
for all $p^*\in E$, we choose $p^*$ with probability proportional to $W^{p^*}_s(1-\epsilon)$ 
(for an appropriate small constant $\epsilon$ that will be determined later) and set
$W'^{p^*}_s=1$, and for any $p\not = p^*$, we set $W'^p_s = 0$. 
After setting all $W'$ variables, for each $p\in E$ and $s\in S$, we set $Z'^p_s = \sum_{t\in E, t\le p} W'^t_s$. 
Finally, for any $q\in Q\setminus S$, $Y'_q = 1$ if and only if $Z'^{c(q)}_s = 1$
for some $s\in S$, such that $s\in D(p)$ (or equivalently $(s,q)\in {\cal C}$).
It is not hard to see that the above rounded integral solution correspond to a feasible bidding 
strategy. In particular, we can implement this bidding
strategy by putting an exact match bid of $b(s)=c(s)$ for any $s\in S$ if $R'_s=1$ (i.e., with probability $R_s$)
and then by putting a broad match bid of $b(s)=p$ for query $s\in S$ if $W'^p_s = 1$ (i.e., with probability
$W^p_s(1-\epsilon)$).


A query $s\in S$ is selected if bid $b(s)$ for this query is at least $c(s)$. 
As a result, a query $s\in S$ is selected with probability
$(1-\epsilon)Z^{c(s)}_s+R_s$. Moreover, for a query $q\in Q\setminus S$ for which $(s,q)\in {\cal C}$ and $(r, q)\in {\cal C}$, 
query $q$ is selected if the bid for either of the queries $s$ or $r$ is at least $c(q)$, i.e., with probability

\begin{eqnarray*}
\Pr[ {\text \ query \ } q {\text \ is\ selected\ }] &= \\
1-(1-(1-\epsilon)Z^{c(q)}_s)(1-(1-\epsilon)Z^{c(q)}_r) &= \\
(1-\epsilon)(Z^{c(q)}_s + Z^{c(q)}_r) - (1-\epsilon)^2Z^{c(q)}_rZ^{c(q)}_s.&
\end{eqnarray*}

Therefore, the expected utility of the solution 
after implementing the integral solution $(V', W', Z', Y')$ 
(or bidding as desribed above) is:
\begin{eqnarray*}
\sum_{s\in S}  ((1-\epsilon)Z^{c(s)}+R_s) w_s + &\\
\sum_{q\in Q\setminus S;r,s\in D(q)} \left( (1-\epsilon)(Z^{c(q)}_s+Z^{c(q)}_r) - 
(1-\epsilon)^2Z^{c(q)}_sZ^{c(q)}_r \right) w_q
\end{eqnarray*}

Next, we derive a lower bound and an upper bound on the probability
that the bid generated as above, wins query $q$. We will show that
\begin{equation}\label{ineq:desired}
Y_q (1-\epsilon)(1-\frac{1}{2}(1-\epsilon))\leq \Pr[q { \text \ is \ selected \ }] \leq (1-\epsilon)2Y_q.
\end{equation}
Consider the following set of inequalities that hold for every
query $q$ that depends on queries $s,r\in D(q)$:

\begin{eqnarray}
\label{IE:probability_q}Z^{c(q)}_sZ^{c(q)}_r \leq \sqrt{Z^{c(q)}_s Z^{c(q)}_r} \leq &\\
\frac{Z^{c(q)}_s+Z^{c(q)}_r}{2} \leq Y_q
\leq Z^{c(q)}_s + Z^{c(q)}_r ,&\nonumber
\end{eqnarray}
The first inequality follows the constraints $0\leq Z^{c(q)}_s,Z^{c(q)}_r \leq 1$
and the second inequality is the arithmetic geometric  mean
inequality. The third inequality follows the summation of the
inequalities $Z^{c(q)}_r \leq Y_q$ and $Z^{c(q)}_s\leq Y_q$ and the last inequality
appears as a constraint in the LP.

The left hand side in Inequality~\ref{ineq:desired} follows
since 

\begin{eqnarray*}
(1-\epsilon)(Z^{c(q)}_s+Z^{c(q)}_r) -(1-\epsilon)Z^{c(q)}_s (1-\epsilon)Z^{c(q)}_r \leq &\\
(1-\epsilon) 2 Y_q - (1-\epsilon)^2 Z^{c(q)}_s Z^{c(q)}_r \leq (1-\epsilon) 2 Y_q
\end{eqnarray*}
And the right hand side follows 
\begin{eqnarray*}
(1-\epsilon)(Z^{c(q)}_s+Z^{c(q)}_r) -(1-\epsilon)^2 Z^{c(q)}_sZ^{c(q)}_r \geq& \\
(1-\epsilon)(Z^{c(q)}_s+Z^{c(q)}_r) -(1-\epsilon)^2 \frac{1}{2}(Z^{c(q)}_s+Z^{c(q)}_r)  \ge &\\
(1-\epsilon)(Z^{c(q)}_s+Z^{c(q)}_r)(1-\frac{1}{2}(1-\epsilon)) \ge &\\
 Y_q(1-\epsilon)(1-\frac{1}{2}(1-\epsilon)).&
\end{eqnarray*}

In the summation that describes the overall utility from the
queries, the probability for selecting query $q$, is multiplied by
both the value and the cost of $q$. Using Inequality
\ref{IE:probability_q}, we get a lower bound on the expected value
from $q$ and an upper bound on the expected cost of $q$.

Let us denote the optimal utility any bidding strategy can
achieve by $U^* = V^* - C^*$, where $V^*$ and $C^*$ are the value and cost part of the objective
utility function, respectively. Let $U^* = U^*_E + U^*_B$ where $U^*_E$
is the utility resulting from the exact match bidding, and
$U^*_B$ is the utility from the broad match bidding. 
Similarly we define $U^*_E = V^*_E - C^*_E$, and
$U^*_B = V^*_B - C^*_B$ (where $V$ and $C$ correspond to the value
and the cost of each part of the solution).
Knowing that for each query $w(q) = v(q) n(q) - c(q) n(q)$, 
the  expected utility of the
above algorithm based on randomized rounding of the LP is at least:
\begin{eqnarray*}
(1-\epsilon)(1-\frac{1}{2}(1-\epsilon))V^*_B - (2-2\epsilon) C^*_B + V^*_E - C^*_E \ge &\\
(1-\epsilon)(1-\frac{1}{2}(1-\epsilon))V^* - \max(1,2-2\epsilon) C^*.&
\end{eqnarray*}

\begin{lemma}
  By setting $\epsilon = 0$ or $\epsilon = 1/2$ in the above algorithm, we get that
  $U^{\mathrm{ALG}}\geq \frac{1}{2}V^* - 2C^*$ or $U^{\mathrm{ALG}}\geq \frac{3}{8}V^* - C^*$, respectively.
\end{lemma}

Given the above lemma, we conclude the following:

\begin{theorem}
For instances of the optimal bidding problem in which $C^*\le {V^*\over 4}$ and each query depend on at most
2 other queries (i.e., $\vert D(q)\vert\le 2$ for each  $q\in Q\setminus S$),
the above randomized algorithm is a $1\over 6$-approximation algorithm.
\end{theorem}

In order to extend the above result for the more general case in which $\vert D(q) \vert \le c'$ for
a constant $c'$, we should add the inequality $Y_q\ge \sum_{s\in D(q)} Z^{c(q)}_s$, and then one
can generalize the above result to the following:
For any constant $c'$, there exist two constants $c$ and $c'$ such that for
instances of the optimal bidding problem in which $C^*\le {V^*\over c}$ and 
$\vert D(q)\vert \le c'$ for each  $q\in Q\setminus S$,  
there exists a constant-factor approximation algorithm.





\section{Budget Constraints}

In this section, we study the problem with an additional budget constraint, i.e, we have
a budget limit $B$ and the
total cost of our bidding strategy should not exceed this limit($B$). Our goal is
to maximize the total value subject to this budget constraint. More formally, the problem is as follows:

\begin{eqnarray} 
\mathrm{Budgeted-IP} \max && \max \sum_{q_i\in Q} X_{q_i} v(q_i)n(q_i)\nonumber \\
\forall (q_j,q_i)\in \mathcal{C} &:& X_{q_i}-X_{q_j}\geq 0 \nonumber\\
 && \sum_{q_i\in Q} X_{q_i} c(q_i)n(q_i) \le B \nonumber\\
 \forall q_i\in Q &:& X_{q_i}\in\{0,1\}\label{IP:budgeted}
\end{eqnarray}

Similar to IP \ref{IP:ineq:pair}, for every query $q$, an integral variable $X_q$ indicates whether
$q$ belongs to the set of queries won by an the optimal solution or
not. There are two difference between the above IP and the IP for query language without the budget constraint: 
One difference is in the objective function in which instead of $w(q_i)$, we have $v(q_i)$, and
the more important difference is the extra budget constraint. 
Because of this extra linear constraint, the following linear programming relaxation of this IP is
not totally unimodular anymore. 

\begin{eqnarray}
\mathrm{Budgeted-LP}&:& \max \sum_{q_i\in Q} X_{q_i} v(q_i)n(q_i)\nonumber\\
 {\text For \ every \ pair \ }(q_j,q_i)\in \mathcal{C} &:& X_{q_i}-X_{q_j}\geq 0\nonumber\\
 && \sum_{q_i\in Q} X_{q_i} c(q_i) n(q_i)\le B \nonumber\\
 \forall q_i\in Q &:& 0\le X_{q_i}\le 1 \label{LP:budgeted}
\end{eqnarray}

\begin{figure}\label{fig:integrality_gap}
\centering
\epsfig{file=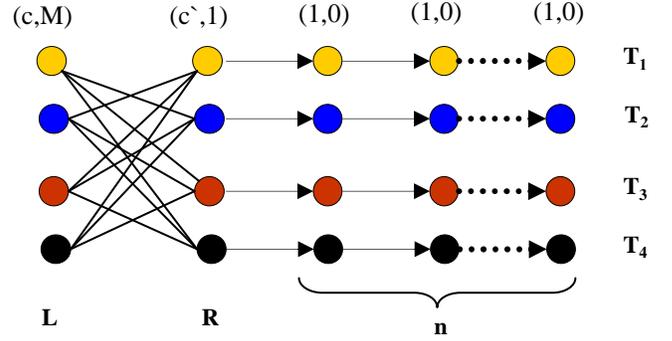}
\caption{An illustration of a bidding problem with a large integrality gap ($k=3$).}
\end{figure}

In fact, we can show that the integrality gap of this LP can be
very large, and thus one cannot round the fractional 
solution of this linear programming relaxation 
to a good integral solution (as we did for IP ~\ref{IP:ineq:pair}).

\begin{lemma}
The integrality gap of linear programming relaxation ~\ref{LP:budgeted}
can be arbitrarily large.
\end{lemma}
\begin{proof}
Consider the following instance of the bidding problem with a budget constraint
in the query language:
Let the set of queries $Q=R\cup L\cup T$ where 
$R = \{r_1, r_2, \ldots, r_{k+1}\}$, $L = \{l_1, l_2, \ldots, l_{k+1}\}$, and
$T = T_1\cup T_2\cup \cdots \cup T_{k+1}$, where $T_i = \{t_{i1}, t_{i2}, \ldots, t_{in}\}$. 
For any query $q\in L$, we have $c(q) = c$ and $v(q) = M$, and for $q\in R$, 
$c(q) = c'$ and $v(q) = 1$. We assume that $c>>c'>>n>>1$. 
For any $q\in T$, $v(q) = 0$, and $c(q) = 1$. 
Also, let the query dependency structure be:
\begin{eqnarray*}
{\cal C} = \{(r_i, l_j)\vert 1\le i,j\le k+1, i\not = j\} \cup &\\
\{(r_i, t_{i1}\vert 1\le i\le k+1\}
\cup &\\
\{(t_{ij}, t_{i(j+1)}\vert 1\le i\le k+1, 1\le j\le n-1\}.
\end{eqnarray*}

It is easy to check that $\cal C$ matches the cost and query definitions.
We also set the budget $B= c + kc'+n$.  In this instance of the bidding problem, 
the optimal integer solution can pick at most one node from the
set $L$ (given the budget constraint of $B$ and the fact that $c>>c'>>n>>1$), 
and thus  the optimal integer solution is $M+k$.

On the other hand, as a fractional solution, we can set $x= {(c+kc'+n)\over ((k+1)c
+(k+1)c' + n)}$. The value of this fractional solution is $[(k+1)M + k + 1]{(c+kc'+n)\over((k+1)c +(k+1)c' + n)}$
which is approximately $(k+1) M$. As a result, the ratio between the optimal fraction
solution and the integral solution can be as large as $k$.
This proves that the integrality gap of the LP is arbitrarily large.
\end{proof}

In fact, not only the LP for the budgeted problem 
is not integral, but also the optimal bidding problem with a budget constraint
is an NP-hard problem even in the query language. 
The NP-hardness follows from the fact that this problem is harder than the knapsack problem.
In fact, the knapsack problem is a special case of this problem in which 
the set of queries are only the keywords and $\cal C=\emptyset$.
Despite the large integrality gap of the above LP and NP-hardness of the problem, 
in the following we show 
how one can use a certain set of optimal solutions of this LP to implement
two budget constrained ad campaigns for an ad auction with broad-match that achieves
the optimal fractional solution of the above LP. 
We show this fact by proving the existence of optimal solutions for the LP with 
certain structural properties. The key structural lemma is the following:

\begin{lemma}\label{thm:structure}
The Linear Program~\ref{LP:budgeted} has at least one optimal solution $X^*$ for which there
exists a value $X$ such that: for each query $q$, $X^*_q \in \{0,1,X\}$.
Moreover, this optimal solution can be found in polynomial time. 
\end{lemma}
\begin{proof}
We prove this fact by showing that an optimal {\em basic feasible
solution} of the LP satisfies the desired properties. From  standard linear programming theory, we know
that such a basic solution exists, and can be found in polynomial time.
Consider an optimal basic feasible solution $X^*$ of the LP~\ref{LP:budgeted}.
Since the LP has $\vert Q\vert$ variables, a basic feasible
solution can be uniquely characterized by 
$\vert Q\vert$ tight inequalities. In other words, there
is a set $P$ of $\vert Q\vert$ independent linear equations among the linear constraints
that characterize $X^*$. Let $P_1\subset P$ be a set of these 
$\vert Q\vert$ linear equations of the form $X^*_{q_i} = 0$ or $X^*_{q_i} = 1$.
Each linear equation in $P_1$ corresponds to an integral variable $X^*_{q_i}$. 
Let $S_1$ be the set of queries $q_i$ corresponding to these integral variables $X^*_{q_i}$. 
Also let $P_2\subset P$ be a set of these 
$\vert Q\vert$ linear equations of the form $X^*_{q_i} = X^*_{q_j}$.
We construct a graph $G(X^*, P_2)$ whose vertex set $V(G)$ is
the set of variables $X^*_{q_i}$ for all queries $q_i\in Q$ 
as follows: we put an edge between a vertex $X^*_{q_i}$ 
to a vertex $X^*_{q_j}$ if and only if $X^*_{q_i} = X^*_{q_j}$
is a linear equation in $P_2$. Thus, if there is
path between two nodes $X^*_{q_i}$ and $X^*_{q_j}$ in $G$, 
we have $X^*_{q_i} = X^*_{q_j}$. Let the connected
components of $G$ be $G_1, G_2, \ldots, G_t$. As a
result, for any two variables $X^*_{q_i}$ and $X^*_{q_j}$ in the 
same connected component $G_p$ for
$1\le p\le t$, we have $X^*_{q_i} = X^*_{q_j}$ (since there is a path
between any two nodes in the same connected component.)
We say that a connected component $G_p$ is a {\em bad} component
if none of the nodes in $G_p$ are in $S_1$. Otherwise, we say that
$G_p$ is a {\em good component}.  
In the following, we show that the number of
bad components is at most one, and this will prove the 
lemma. In order to prove this claim, we need to prove the following lemma:

\begin{lemma}\label{lem:nocycle}
Graph $G$ as defined above does not have any cycle.
\end{lemma}
\begin{proof}
For contradiction, assume that there exists a cycle $X^*_1$, 
$X^*_2$, $\cdots$, $X^*_v$ in graph $G$. Thus, 
all linear equations $X^*_i = X^*_{i+1}$ for $1\le i\le v-1$, and
$X^*_v = X^*_1$ are in $P$. But these $v$ equations are not
independent, and this cycle contradicts the fact that 
$P$ is a set of independent linear equations. 
\end{proof}

Lemma~\ref{lem:nocycle} proves that graph $G$ is a forest, and
thus connected components $G_i$ for $1\le i\le p$ are all trees, 
and each has $\vert V(G_i)\vert-1$ edges. Now, 
we observe that for any good component $G_i$,  there exists exactly 
one equation in $P_1$. For any good component $G_i$, 
there are $\vert V(G_i)\vert-1$ equations corresponding to
edges of $G_i$, and at least one equation in $P_1$. If
there two such equations in $P_1$ for nodes of $G_i$, then the union of
these two equations and equations corresponding to edges
of $G_i$ form $\vert V(G_i)\vert+1$ equations
in $P$ all defined on only variable on $V(G_i)$.
As a result, these equations cannot be independent, which contradicts with
the fact that $P$ is a set of independent equations. 
As a result, each good component $G_i$ corresponds
to exactly $\vert V(G_i)\vert$ equations in $P$. 

Let $Y$ be the set of variables $X^*_{q_i}$ in good connected components
of $G(X^*, P_2)$. Thus, $Y$ corresponds to a set of all integral variables $X^*_{q_i}$.
The above discussion implies that there are 
exactly $\vert Y\vert$ equations in $P$ characterizing all variables in $Y$.
As a result, there are $\vert P\vert-\vert Y\vert$ equations 
uniquely identifying all the (fractional) variables in $Q\backslash Y$. Noting that
$\vert P\vert = \vert Q\vert$, we conclude
that $\vert Q\backslash Y\vert$ equations 
uniquely identify all the (fractional) variables in $Q\backslash Y$, 
and none of these equations are of the form $X^*_{q_i} =0$ or 
$X^*_{q_i} =1$. At most one of these equations correspond to a tight
budget constraint ($\sum_{q_i\in Q} X^*_{q_i} c(q_i) n(q_i) = B$), 
and thus we have at least $\vert Q\backslash Y\vert -1$ equations
of the form $X^*_{q_i} = X^*_{q_j}$  on variables $Q\backslash Y$ 
in bad components. Since there is no cycle in graph $G$ and 
vertices in $Q\backslash Y$ have at least $\vert Q\backslash Y\vert -1$ edges
amongst them, they should all belong to the same connected component (i.e,
the only one bad connected component),  
and thus have the same value $X^*_{q_i} = X$. This completes the 
proof of the lemma.
\end{proof}

Using Lemma \ref{thm:structure}, we can show that in the 
query language model, one can implement the optimal fractional solution using two 
budget-constrained ad campaigns. To formally show this, we assume that queries 
arrive at the same rate, and by putting a budget constraint on an ad campaign
(that includes a set of queries), the budget on all queries is consumed
at the same rate until it gets used completely.

\begin{theorem}
In the query language model, there exists a polynomial-time 
algorithm that computes two budget-constrained ad campaigns
that implement an optimal bidding strategy achieving the maximum  value
for the advertiser given a budget constraint.
\end{theorem}
\begin{proof}
The polynomial-time algorithm is as follows:

\begin{enumerate}
\item Solve LP~\ref{LP:budgeted} and compute an optimal solution $X^*$
such that $X^*_{q_i} \in \{0,1,X\}$ for all queries $q_i$.
\item Let $S_0$ and $S_1$ be the sets of queries with the corresponding 
 integral variables $X^*_{q_i}=0$ and  $X^*_{q_i}=1$, respectively. 
\item Let $B_1 = \sum_{q_i\in S_1} X^*_{q_i} c(q_i) n(q_i)$. 

\item Run the following two ad campaigns:

\begin{enumerate}
  \item A campaign with budget $B_1$ on queries in $S_1$.
  \item A campaign with budget $B-B_1$ on queries in $Q\backslash (S_0\cup S_1)$.
\end{enumerate}

\end{enumerate}

Lemma \ref{thm:structure} shows that the first step of this algorithm
can be done in polynomial time.  
To show the correctness of the algorithm, note
that assuming that queries arrive at the same rate, 
and based on the definition of the budget-constrained ad 
campaign, the value from these two ad campaigns is:

$$\sum_{q\in S_1} v(q) n(q) + {B-B_1\over B'} \sum_{q\in  Q\backslash (S_0\cup S_1)} v(q) n(q),$$
where $B'= \sum_{q\in  Q\backslash (S_0\cup S_1)} c(q) n(q)$.
Note that $X^*_{q_i} = 1$ for $q_i\in S_1$, we have 
$B_1 = \sum_{q_i\in S_1} X^*_{q_i} c(q_i) n(q_i) =\sum_{q_i\in S_1}  c(q_i) n(q_i)$, 
and thus $B_1 + \sum_{q\in  Q\backslash (S_0\cup S_1)} X c(q) n(q) = B$, therefore, 
$$X= {B-B_1\over \sum_{q\in  Q\backslash (S_0\cup S_1)} c(q) n(q)}={B-B_1\over B'}.$$  Thus, the value of
the optimal solution is:

\begin{eqnarray*}
\sum_{q\in S_1} v(q)n(q) + \sum_{q\in  Q\backslash (S_0\cup S_1)} X v(q) n(q) = &\\
\sum_{q\in S_1} v(q)n(q) + {B-B_1\over B'} \sum_{q\in  Q\backslash (S_0\cup S_1)} v(q) n(q).
\end{eqnarray*}
Hence, the total value from these two campaigns is the same as the optimal fractional solution, 
as desired.
\end{proof}

Finally, we observe that the optimal bidding problem with a budget constraint
is APX-hard for the keyword language.

\begin{theorem}
The optimal bidding problem with a budget constraint is APX-hard for the keyword language. Moreover, 
this problem is not approximable within a factor better than a multiplicative factor $1-{1\over e}$. 
\end{theorem}
\begin{proof}
We give a simple reduction from the maximum coverage problem. In an instance 
of the maximum coverage problem, we are given a family of subset $S_1, \ldots, S_k\subset V$,
and a value $w(e)$ for each element $e\in V$. The goal is to find a 
family of $k$ subsets $S_{a_1}, \ldots, S_{a_k}$ that maximizes
$\sum_{e\in \cup_{i\le k} S_{a_i}}$. Given an instance of the 
maximum coverage problem, we define an instance of optimal bidding
problem as follows: for each subset $S_i$ ($1\le i\le k$) in the maximum coverage problem, we
put a keyword $s_{S_i}$ in the set $S$ of keywords in the optimal bidding problem. 
The cost $c(s)$ of each keyword in $S$ is 1 and its value is zero. 
We also put a query $q_e\in Q$ corresponding to each element $e\in V$ of the maximum
coverage problem. The value of query $q_e\in Q\setminus S$ is one and the cost 
of each query is zero. Moreover, we say that a query $q_e$ corresponding to an element
$e$ can be broadly matched with any query $s_{S_i}\in S$ corresponding to
any subset $S_i$ of the maximum coverage problem that includes the element $e$, i.e.,
$e\in S_i$. Finally, we set the budget constraint $B$ to $k$. 
It is not hard to see that the maximum value bidding strategy on keywords
in this instance with the total cost at most $B$ corresponds
to the maximum value set-coverage of at most $k$ sets in the original problem.
This reduction implies that our problem is not approximable
within a factor better than $1-{1\over e}$ unless P=NP, since
the maximum coverage problem is NP-hard to approximate within a factor
better than $1-{1\over e}$~\cite{F98}. 
\end{proof}

\section{An Experimental Study}\label{sec:simulation}
In this section, we report results from an experimental study to address
how much an ad campaign loses by using solely broad match rather than
a combination
of exact and broad match types.
Our simulation is composed of 30 keywords, where we consider all pairs of
keywords as the set of possible queries. Therefore, while there are ``only'' 30
keywords an advertiser who is interested in managing all possible
queries will have $435$ keywords to consider, which is tedious
for small advertisers.
Most advertisers will prefer a campaign with a small set of keywords
which they can
easily track and evaluate, which in our simulation is represented by
the core 30 keywords. The setup is very simple. All
queries have the same cost. The net value of a query
is determined as follows. The value of a keyword is drawn from a
standard normal distribution. The net value of a query is either
(1) the average net value of its keywords, or (2) the max value among
its keywords, or (3) the min value among its keywords; the precise
net value of a query is decided according to $1-3$ uniformly at random.
This setting is loosely motivated by the intuition that some queries just
average the keywords in the query (like ``Canon or Nikon''), some are valued
as the best  among the keywords (like ``Canon DSLR''), and some valued
as the worst among the keywords (like ``Canon calculator'').

Running the simulation 15 times, we obtain that the average value
obtained by solving the integer linear program while allowing both
exact match and broad
match was $120.9$ while allowing only broad match was $119.2$.
Furthermore we obtain that the maximum ratio between the two was less
than $4$ percent. This simple simulation supports our initial hope
that not using
exact match may be a realistic assumption for some advertisers, in
particular, small to medium
advertisers. We must remark that a more detailed experimental study is needed
to be more conclusive. Our hope is that our LP-based algorithms can
indeed be run with
reasonably sized problems for this purpose.

\ignore{
A similar simulation for $20$ keywords in which the advertiser can
choose an exact match for every query yields that the average
performance is better by $21$ percent (note that the number of
queries to manage is now 190 instead of 20 for a broad match campaign)
with standard deviation of $14$ percent.  Therefore, an advertiser
who can manage and track a large number of keywords and queries can
boost its revenue considerably, however for many small advertisers
boosting their revenue by $20$ percent will not return their
investment on managing complex advertising campaigns.}

\section{Concluding Remarks}
Our work initiates the study of the bid optimization problem for advertisers in presence
of a common feature in sponsored search, ie., the broad match type.
The central technical issue is that choosing to bid on a keyphrase may yield positive 
profit from some queries, but may commit one to implicitly bid on queries in which the profit
may be small or even negative. We propose LP-based polynomial-time algorithms for
this problem which is optimal under the query language model, and is an approximation
in the keyword language model for certain cases while it is NP-Hard to even approximate 
the optimal solution to any factor, in general. 

Our work leaves open several research problems. A technical problem is to extend the results here to
the multi-slot case. More precisely, given a ``landscape'' that is a function of bids and 
gives estimated clicks and cost, obtain the profit-maximizing bidding strategy.
A conceptual problem is to determine a suitable approach for broad match auctions where 
search engines are able to provide faithful estimates for clicks and cost associated with
the broad matches, so advertisers can bid accurately. This involves averaging over 
many related queries. A principled approach to formulating this notion  will be of great
interest.

Finally, a specific technical question that remains open in this paper is the approximability
of the budget-constrained version of the optimal bidding problem in the keyword language.
An interesting aspect of this problem is the following relation  
to submodular optimization.
Given a subset $T\subset S$ of keywords on which 
we can bid, let us denote the total cost and value of the queries 
that we win as a result of bidding on keywords in $T$ by $C(T)$
and $V(T)$. One can check that  both the value 
and cost functions $V, C: 2^S\rightarrow R$ are set-cover-type set functions,
and thus they are monotone and submodular.~\footnote{A set function $f: 2^S\rightarrow R$ 
is submodular if and only for any two subsets $A$ and $B$, 
$f(A) + f(B) \ge f(A\cap B) + f(A\cup B)$.} 
The optimal bidding problem with a budget constraint 
in the query language is, therefore, to find a subset $T\subset S$ of
keywords that maximizes the submodular function $V(T)$ subject to
the submodular constraint $C(T)\le B$. Constant-factor approximation algorithms 
are known for maximizing a general monotone submodular function subject to a knapsack (modular)
constraint~\cite{Sviri}, but maximizing submodular functions subject to 
{\em a submodular constraint} (as in our case) is an open question.

\bibliographystyle{abbrv}

\end{document}